\documentclass[aps]{revtex4}
\usepackage{epsfig}

\begin{document}

\title{Better detection of Multipartite Bound Entanglement
with Three-Setting Bell Inequalities}

\author{Dagomir Kaszlikowski}
\affiliation{Department of Physics, National University of
Singapore, 10 Kent Ridge Crescent, Singapore 119260}
\author{L. C. Kwek}
\affiliation{Department of Physics, National University of
Singapore, 10 Kent Ridge Crescent, Singapore 119260}
\affiliation{National Institute of Education, Nanyang
Technological University, 1 Nanyang Walk, Singapore 639798}
\author{Jing Ling Chen, C. H. Oh}
\affiliation{Department of Physics, National University of
Singapore, 10 Kent Ridge Crescent, Singapore 119260}
\begin{abstract}
It was shown in Phys. Rev. Lett. {\bf 87}, 230402 (2001) that N
($N\geq 4$) qubits described by a certain one parameter family
${\cal F}$ of bound entangled states violate Mermin-Klyshko
inequality for $N\geq 8$. In this paper we prove that the states
from the family ${\cal F}$ violate Bell inequalities derived in
Phys. Rev. A{\bf 56}, R1682 (1997), in which each observer
measures three non-commuting sets of orthogonal projectors, for
$N\geq 7$. We also derive a simple one parameter family of
entanglement witnesses that detect entanglement for all the
states belonging to ${\cal F}$. It is possible that these new
entanglement witnesses could be generated by some Bell
inequalities.
\end{abstract}
\maketitle

\section{Introduction}

Entanglement has been a very useful resource for information
processing. Some form of entangled pure states were required for
achieving the desired results in quantum applications like
quantum dense coding Ref. \cite{dense}, quantum teleportation
Ref. \cite{tele} and so forth. In
practical situations, we do not have pure entangled states due to
noise from the environment.  Consequently, there is a need to
distill using only local operations and classical communications
(LOCC) some amount of pure state entanglement.  While it is true
that all pure states with non-zero entanglement can be
transformed by LOCC with some probability to maximally entangled
states, there is no general criterion for determining if a mixed
entangled state can be distiled. Surprisingly, it was shown by
Horodecki \cite{horo1} that there are mixed states which cannot be
distiled despite being entangled. Such states are now known as
bound entangled states. The discovery of bound entanglement
Ref. \cite{horo1} immediately posed two important questions:
Firstly, is bound entanglement useful in quantum information?
Secondly, do bound entangled states admit local, realistic
description?.

The answer to the first question was first given in
Ref. \cite{HORODECKI-ACTIV} where it was shown that one can increase
the fidelity of the so-called conclusive teleportation with the
aid of bound entanglement. Moreover, the increase in fidelity
beyond some threshold cannot be achieved without the use of bound
entanglement. Therefore, bound entanglement can be a valuable
resource in quantum information processing.

However, there is only a partial answer to the second question.
On the one hand it was shown in Ref. \cite{DUR} that some bound
entangled $N$-qubit states from the family ${\cal F}$ do not admit
local, realistic description for $N\geq 8$ and yet on the other
hand it was shown numerically that there exists a bound entangled
state Ref. \cite{BENNETT-UPB} in $3\otimes 3$ Hilbert space which
does not violate local realism Ref. \cite{KASZ-BOUND}.

The main difficulty in answering the second question lies with the
fact that we do not have analytical tools (usually, Bell
inequalities) that provide necessary and sufficient conditions for
the existence of a local, realistic description of a bi- or
multi-partite quantum systems except for the two-qubit case. In
the two-qubit case, one has the Clauser-Horne inequalities but
unfortunately no bound entangled states exist. Recently a set of
Bell inequalities for higher dimensional bipartite systems
Ref. \cite{KASZ-CH, POPESCU} as well as for multipartite qubit
systems Ref. \cite{BRUKNER, WERNER} were discovered. However, these
inequalities have not been proven to be sufficient conditions for
the existence of a local realistic description. Indeed, it was
shown that the inequalities derived in Ref. \cite{BRUKNER} are not
violated for certain pure non-maximally entangled states
Ref. \cite{GISIN-WIESNIAK}. This means that these inequalities are
only necessary condition for a local realistic description. The
only available method currently known uses a linear programming
algorithm to extract necessary and sufficient conditions for
local realism. However, this algorithm is computationally
inefficient.

Note that the most versatile tool possible for investigating the
existence of a local realistic description of quantum
correlations, namely Bell inequalities, has a serious
drawback. The reason is that Bell inequalities are generally
derived for the situation in which each observer measures two
non-commuting observables (i.e. two settings of measuring
apparatus). Therefore, even if we have a set of necessary and
sufficient Bell inequalities for two settings of measuring
apparatus and if these inequalities are not violated for a
given state, it may happen that there exists another set of Bell
inequalities utilizing more than two settings of measuring
apparatus that could be violated by the state. Thus, although
there is strong numerical evidence that this is not the case for
the two-qubit systems Ref. \cite{BATURO}, it was shown in Ref.
\cite{ZUK-KASZ} that one can derive Bell inequalities, in which
each observer uses three settings of the measuring apparatus, to
obtain a stronger violation of local realism than the inequalities
for two settings.

In this paper, we will show that the three-setting Bell
inequalities derived in Ref.\cite{ZUK-KASZ} are violated by all
the states from the family ${\cal F}$ for $N\geq 7$. Note that
violation of the two-setting Mermin-Klyshko Ref. \cite{MERMIN}
inequalities requires
an additional qubit. From the experimental perspective, the
reduction of even one qubit has a distinct advantage in which one
knows that it is always easier to produce and control fewer
qubits. Moreover, the lowering of the number of qubits using
three-setting Bell inequalities may imply that perhaps one can
violate local realism even with four qubits using states from the
above family and stronger Bell inequalities, perhaps with four or
more settings. The family ${\cal F}$ is also the first example of
entangled states for which $d$-setting ($d>2$) Bell inequalities
reveal entanglement whereas two-setting ones do not.

We have also derived, based on the structure of the one parameter
family of entanglement witnesses generated by the three-setting
Bell inequalities, a one parameter family of entanglement
witnesses that can detect entanglement for all the bound
entangled states states from the family ${\cal F}$ for $N\geq 4$.

\section{Three-setting Bell inequality}

In Ref. \cite{ZUK-KASZ}, a series of Bell inequalities for $N$
entangled qubits, in which each observer measures three
non-commuting observables, was derived. It was shown that the
violation of these inequalities by the GHZ state $|\psi\rangle$
of the form
\begin{eqnarray}
&&|\psi\rangle = {1\over\sqrt 2}(|0\rangle^{\otimes N} +
|1\rangle^{\otimes N}), \label{GHZ}
\end{eqnarray}
where the two states $|0\rangle, |1\rangle$ form an orthogonal
basis in the Hilbert space of each qubit is stronger than
Mermin-Klyshko Ref. \cite{MERMIN} inequalities for $N\geq 4$.

Let us first obtain the Bell operator generated by the
inequalities in Ref. \cite{ZUK-KASZ}. The inequalities have the
following structure
\begin{equation}
-2^{N-1}\sqrt 3\leq
\sum_{k_1,k_2,\dots,k_N=1}^{3}c_{k_1k_2\dots_N}E_{k_1k_2\dots
k_N}\leq 2^{N-1}\sqrt 3. \label{inequality}
\end{equation}
The coefficients of the inequality read $c_{k_1k_2\dots
k_N}=cos(\phi^1_{k_1}+\phi^2_{k_2}+\dots +\phi^N_{k_N})$, where
$\phi^1_1={\pi\over 6}, \phi^1_2={\pi\over 2},
\phi^1_3={5\pi\over 6}$ and $\phi^n_1=0, \phi^n_2={\pi\over
3},\phi^n_3={2\pi\over 3}$ for $n=2,3,\dots,N$ are the phases
associated with three sets of two orthogonal projectors
corresponding to the three settings of the measuring apparatus
measured by each observer. Here, the superscripts denote the
observers and the subscripts enumerate the set of the projectors.
We assume that each set of projectors is obtained from the set
$P^{k}_0(0)=|0\rangle\langle 0|, P^{k}_0(1)= |1\rangle\langle 1|$
by the rotation using the unitary operators $U(\phi^{k}_1),
U(\phi^{k}_2, U(\phi^{k}_3))$ of the form
\begin{equation}
U(\phi^{k}_l)={1\over\sqrt 2}\left(\begin{array}{cc}
1 & 1\\
\exp(i\phi^k_l) & -\exp(i\phi^k_l)\end{array}\right),
\end{equation}
where $l=1,2,3$. Therefore, the $k$-th observer in the $l$-th
experiment measures two projectors
$P^{k}_l(0)=U(\phi^{k}_l)|0\rangle\langle 0|U(-\phi^{k}_l),
P^{k}_l(1)=U(\phi^{k}_l)|1\rangle\langle 1|U(-\phi^{k}_l)$.
Corresponding to the result of the measurement of the projector
$P^{k}_l(0)$, we ascribe the value $-1$ whereas we ascribe the
value $+1$ corresponding to the result of the measurement of the
projector $P^{k}_l(1)$. $E_{k_1k_2\dots k_N}$ is the correlation
function defined in the standard way
\begin{eqnarray}
&&E_{k_1k_2\dots k_N} = \nonumber\\
&&Tr(\rho \sum_{l_1 l_2\dots l_N=1}^{2}(-1)^{l_1+l_2+\dots l_N}
P^{1}_{k_1}(l_1)\otimes P^{2}_{k_2}(l_2)\otimes\dots\otimes
P^{N}_{k_N}(l_N)),
\end{eqnarray}
where $\rho$ is an arbitrary quantum state.

The Bell operator $B_N$ for the inequality reads
\begin{eqnarray}
&&B_N = \sum_{k_1,k_2,\dots,k_N=1}^{3}c_{k_1k_2\dots_N}\nonumber\\
&&\sum_{l_1 l_2\dots l_N=1}^{2}(-1)^{l_1+l_2+\dots l_N}
P^{1}_{k_1}(l_1)\otimes P^{2}_{k_2}(l_2)\otimes\dots\otimes
P^{N}_{k_N}(l_N).
\end{eqnarray}
It is convenient to write this operator in the matrix form. Using
the basis defined by the vectors $|00\dots 0\rangle, |00\dots
1\rangle, |00\dots 10\rangle, |00\dots 11\rangle, \dots, |11\dots
10\rangle, |11\dots 11\rangle$, this operator reads
\begin{equation}
B_N = \left(\begin{array}{ccccc}
0 & 0 & \dots & 0 & {(-3)^N\over 2}\\
0 & 0 & \dots & 0 & 0\\
\vdots & \vdots & \vdots & \vdots & \vdots\\
{(-3)^N\over 2} & 0 & \dots & 0 & 0 \end{array}\right). \label{op}
\end{equation}
Note that only the matrix elements $\langle 00\dots
0|B_{N}|11\dots 1\rangle$ and $\langle 11\dots 1|B_{N}|00\dots
0\rangle$ are non zero.

It is easy to see that for the GHZ state (\ref{GHZ})
$Tr(B_N|\psi\rangle\langle\psi |)= {(-3)^N\over 2}$, i.e., one
has a violation of the inequality (\ref{inequality}).  For $N\geq
4$, this violation is stronger than the violation obtained
through the Mermin-Klyshko inequality.

It is instructive to note that the Bell operator $B_N(\alpha_N)$,
of the form
\begin{eqnarray}
&&B_N(\alpha_N) = U({\alpha_N\over N})^{\otimes
N}B_N U({-\alpha_N\over N})^{\otimes N},
\label{rotop}
\end{eqnarray}
where $U({\alpha_N\over N}) = |0\rangle\langle 0| +
\exp(i{\alpha_N\over N})|1\rangle\langle 1|$, is optimal for the
violation of the three-setting Bell inequality for the state
$|\psi(\alpha_N) = {1\over \sqrt 2}(|0\rangle^{\otimes N} +
\exp(i\alpha_N)|1\rangle^{\otimes N})$. To see this, one observes
that the state $|\psi(\alpha_N)\rangle$ can be obtained through
local rotation by each observer using the unitary transformation
$U({\alpha_N\over N})$ on their portion of the state
$|\psi\rangle$.

\section{Violation of Local Realism}

It was shown in Ref. \cite{DUR} that the following one-parameter
family ${\cal F}$ of $N$-qubit states
\begin{eqnarray}
&&\rho_N(\alpha_N) = {1\over (N+1)}
(|\psi(\alpha_N)\rangle\langle\psi(\alpha_N)| + {1\over
2}\sum_{k=1}^{N}(P_k+\bar{P}_k)),
\end{eqnarray}
where $P_k$ is a projector on the state $|00\dots 010 \dots
00\rangle$ with $1$ being on the $k$-th position and $\bar{P}_k$
is a projector on the state $|11\dots 101 \dots 11\rangle$ with
$0$ being on the $k$-th position, is a family of states that are
entangled but which cannot be distiled if $N\geq 4$. By
distillation we understand, following Dur, the impossibility of
extracting any pure entangled state from the states belonging to
the family by means of LOCC. Dur also showed that for $\alpha_N =
{\pi\over (4(N-1))}$, these states violate Mermin-Klyshko
inequality for $N\geq 8$. The latter result means that they do
not admit a local realistic description. As noted earlier, this
is the first known example of bound entangled states violating
local realism.

Let us now apply the rotated Bell operator in Eq.(\ref{rotop}) to
$\rho_{N}(\alpha_N)$. A straightforward computation gives
\begin{eqnarray}
&&Tr(B_N(\alpha_N)\rho_N(\alpha_N)) = {(-3)^N\over 2(N+1)}.
\end{eqnarray}
To obtain a violation of the inequality (\ref{inequality}) we must
have $|{(-3)^N\over 2(N+1)}| > 2^{N-1}\sqrt 3$ which happens for
$N\geq 7$.

Therefore, we have shown that local realism is violated for the
bound entangled states from the given family if $N\geq 7$
regardless of the parameter $\alpha_N$.

Moreover, the strength of violation of local realism for the
three-setting Bell inequalities presented here is greater than for
Mermin-Klyshko inequalities. In our case, the strength of
violation is defined as the minimal amount $V_N$ ($0\leq V_N\leq
1$) of pure noise $\rho^{noise}_N = {2^{-N}}I\otimes
I\otimes\dots\otimes I$ one has to add to the state
$\rho_N(\alpha_N)$ so that the resulting state
$\sigma_N(\alpha_N) = (1-V_N)\rho_N(\alpha_N) + V_N\rho_{noise}$
does not admit a local realistic description. We have
$Tr(B_{N}(\alpha_N)\sigma_N(\alpha)) = (1-V_N){(-3)^N\over
2(N+1)} $. Thus, the violation occurs for
$V_N>{3^N-2^N(N+1)\sqrt3\over 3^N}$. For instance for $N=7
\mbox{\rm or ~} 8$ one must add around $19\% \mbox{\rm or ~} 39\%$
of noise respectively in order to allow the state
$\rho_{7(\mbox{or ~} 8)}(\alpha_{7( \mbox{\rm or ~ }8)})$ to have
a local realistic description when using the three-setting
inequality. It is important to note that the amount of noise
necessary for a local realistic description in the case of
Mermin-Klyshko inequalities for $N=8$ (for $N=7$ there is no
violation) is $20\%$.

\begin{center}
\begin{figure}
\psfig{file=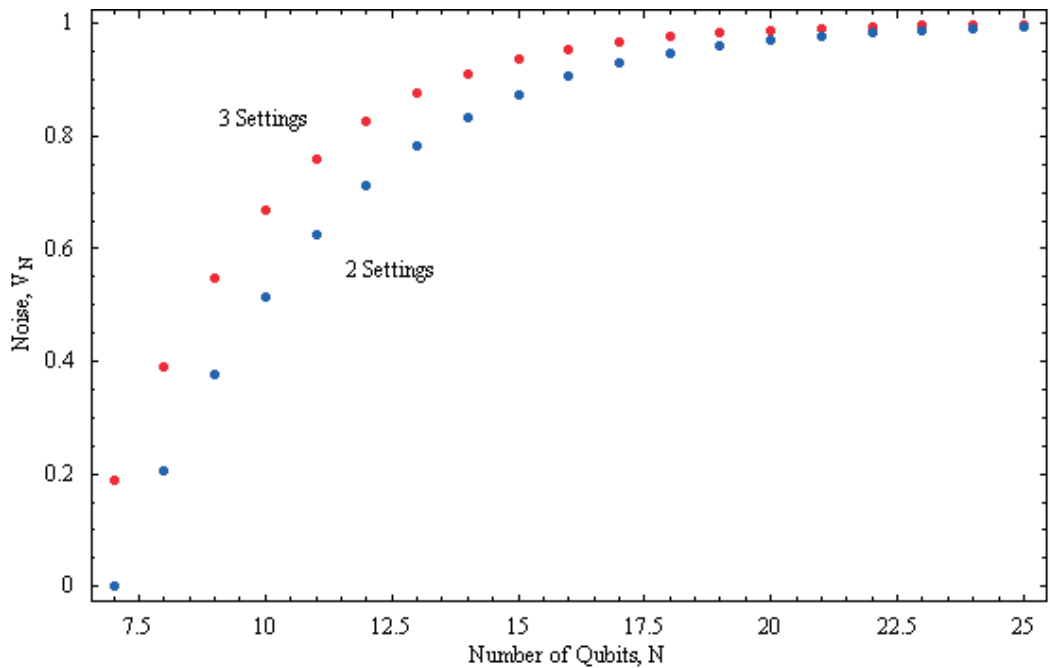,width=0.8\linewidth} \caption{Function of
the amount of noise, $V_N$, as a variation of the number of
qubits, $N$, for the three-setting and two-setting inequalities.}
\label{fig1}
\end{figure}
\end{center}

It is still an open question whether there exist stronger Bell
inequalities that are violated by {\it all} the states from the
family, i.e. for every $N\geq 4$ and for any choice of the
parameter $\alpha_N$. We next provide some numerical evidences
that for $N=4$ one cannot violate local realism if each observer
is allowed to measure two or three sets of projectors.

As shown in Ref. \cite{BATURO, KASZ-PHD, KASZ-ZEIL} one can check,
using linear programming algorithm, if a given quantum state
admits local realistic description. Due to the computational
complexity of linear programming one must resort to numerical
methods. Let us briefly describe the idea of testing local
realism by means of linear programming Ref. \cite{BATURO, KASZ-PHD,
KASZ-ZEIL}.

In a Bell experiment involving $N$ observers measuring $M$ sets of
projectors (each set consisting of two orthogonal projectors) on a
$N$-qubit state, the outcome is a set of $M \times 2^N$
probabilities which we can denote as $p(l_1({k_1}) l_2({k_2})
\dots l_N({k_N});\vec{\phi_{k_1}},\vec{\phi_{k_2}},\dots,
\vec{\phi_{k_N}})$ where $k_i=1,2,\dots, M$ ($i=1,2,\dots,N$).
These labels, $k_i=1,2,\dots, M$ ($i=1,2,\dots,N$), tell us which
set of projectors is measured by $i$-th observer. The index
$l_i({k_i})=0,1$ denotes the outcome of the measurement for $i$-th
observer should he measure the $k_i$-th set of projectors. The
vector $\vec{\phi_{k_i}}$ is a set of real parameters (the vector
notation has only a symbolic meaning) defining the $k_i$-th set
of projectors (for instance, $\vec{\phi_{k_i}}$ has three
components if each observer applies $SU(2)$ transformation to his
qubit). The correlations between the outcomes of the local
measurements performed by the observers are the only information
available according to quantum mechanics. Nevertheless, a local
realistic theory tries to go further. In a local realistic
theory, the basic assumption is that each particle carries a
probabilistic or deterministic set of instructions regarding how
to respond to all possible local measurements that it might be
subjected to. Therefore local realism assumes the existence of
non-negative joint probabilities (summing up to unity) involving
all possible observations from which it should be possible to
obtain all the quantum predictions as marginals. Let us denote
these hypothetical probabilities by $p^{HV}(l_1(1), l_1(2), \dots
,l_1(M);l_2(1), l_2(2), \dots, l_2(M);\dots;l_N(1), l_N(2), \dots
,l_N(M))$. The local realistic probabilities for experimentally
observed events are the marginals
\begin{eqnarray}
&&p^{HV}(l_1(k_1), l_2(k_2), \dots ,l_N(k_N))\nonumber \\
&&\sum_{l_1(m_1)\neq l_1(k_1)}\sum_{l_2(m_2)\neq l_2(k_2)}\dots \nonumber \\
&&\sum_{l_N(m_N)\neq
l_N(k_N)} p^{HV}(l_1(1), l_1(2), \dots, l_1(M); l_2(1), l_2(2), \dots,
l_2(M);\dots;l_N(1), l_N(2), \dots, l_N(M)). \label{horror}
\end{eqnarray}
The $M \times 2^N$ equations in (\ref{horror}) form the complete
set of necessary and sufficient conditions for the existence of a
local realistic description of the experiment. Thus, if it is
possible to find such a joint probability distribution so that
\begin{eqnarray}
&&p(l_1({k_1}), l_2({k_2}), \dots,
l_N({k_N});\vec{\phi_{k_1}},\vec{\phi_{k_2}},\dots,
\vec{\phi_{k_N}}) \nonumber \\ &= & p^{HV}(l_1(k_1),
,l_2(k_2),\dots,l_N(k_N))
\label{marg}
\end{eqnarray}
for the given choice of projectors then the quantum probabilities
have a local realistic description.

Replacing $p^{HV}(l_1(k_1),\dots,l_2(k_2),\dots
,l_N(k_N))$ by the right-hand side
of (\ref{horror}) and putting it into (\ref{marg}), we get a set
of $M \times 2^N$ linear equations with $2^{M+N}$ unknowns (joint
probabilities). Therefore, we have more unknowns than equations.
Moreover we have a set of linear constraints on the unknowns
since these unknowns must be non-negative and they must also sum
up to unity.

The linear programming algorithm allows us to check if there is a
solution to the above set of equations (for details see
Ref. \cite{KASZ-BOUND}). However, we must remember that the left-hand
side of the equations depends on the parameters defining the
measured projectors. Therefore even if there is a solution to the
equations for some choice of the projectors, we do not know if
such a solution exists for some other set of the projectors.

We applied the above method to the bound entangled states from the
given family for $N=4$ with $\alpha_N$ being $0, {\pi\over 2},
\pi$ and in which for each choice of $\alpha_N$, each observer can
measure two or three sets of projectors. The sets of projectors
were obtained by the rotation of the projectors $|0\rangle\langle
0|, |1\rangle\langle 1|$ through unitary operators chosen from the
$SU(2)$ group. For each possible case, we picked 1000 randomly
chosen sets of projectors for each observer. We found solutions
to the appropriate set of equations in all cases. This strongly
suggests that one cannot violate local realism with the four
qubit states chosen from the family. Perhaps, one should consider
wider sets of projectors but this computation lies beyond current
capability of the computers at our disposal.

\section{Strong Entanglement Witnesses}

Multipartite entanglement witness $W$ is a hermitian operator
with the property that $Tr(WP_1\otimes P_2\otimes\dots\otimes
P_N)\geq 0$ for any projectors $P_k$ ($k=1,2,\dots,N$) and there
exist some entangled states $\rho_{ent}$ for which
$Tr(W\rho_{ent})<0$. In the latter case, we say that $W$ detects
the entanglement of $\rho_{ent}$.

The family of entanglement witnesses associated with the Bell
inequality (\ref{inequality}) has the form
\begin{eqnarray}
&&W_N(\alpha_N) = 2^{N-1} \sqrt3 I - |B|_{N}(\alpha_N),
\label{witness}
\end{eqnarray}
where $|B|_{N}(\alpha_N)$ is the operator from Eq. (\ref{rotop})
with the matrix elements replaced by their moduli. It is
convenient to ``normalize'' the operators in the above equation
by dividing them with the number $2^{N-1}\sqrt 3$. Therefore we
need only consider the entanglement witnesses of the form
\begin{equation}
W_N = \left(\begin{array}{ccccc}
1 & 0 & \dots & 0 & -{3^N\over 2^N \sqrt3}\\
0 & 1 & \dots & 0 & 0\\
\vdots & \vdots & \vdots & \vdots & \vdots\\
-{3^N\over 2^N \sqrt3} & 0 & \dots & 0 & 1 \end{array}\right).
\end{equation}

It is interesting to find a family of entanglement witnesses
having the similar structure to the above ones but that can
detect entanglement for all the states from the family ${\cal
F}$, i.e., for $N\geq 4$. To this end let us consider a new
family of entanglement witnesses $S_N$ of the form
\begin{equation}
S_N = \left(\begin{array}{ccccc}
1 & 0 & \dots & 0 & -|S_N|\\
0 & 1 & \dots & 0 & 0\\
\vdots & \vdots & \vdots & \vdots & \vdots\\
-|S_N| & 0 & \dots & 0 & 1 \end{array}\right),
\end{equation}
where $S_N$ is a real number.

We note that the operators $S_N$ must first be positive on all the
product projectors $P_1\otimes P_2\otimes\dots\otimes P_N$.
Therefore, we have the following condition for $|S_N|$
\begin{eqnarray}
&&0\leq Tr(S_N P_1\otimes P_2\otimes\dots\otimes P_N) = 1+2|S_N|Re[\prod_{k=1}^N\langle 0|P_k|1\rangle]\nonumber\\
&&=1+2|S_N|Re[\exp(-i\sum_{k=1}^N\phi_k)\prod_{k=1}^N\sin\theta_k\cos\theta_k]=
1+2^{1-N}|S_N|\cos(\sum_{k=1}^N\phi_k)\prod_{k=1}^N\sin2\theta_k,
\end{eqnarray}
where we used the fact that every pure state of a qubit can be written as
$\cos(\theta)|0\rangle+\exp(i\phi)\sin\theta|1\rangle$. The above equation is
positive for arbitrary product projectors if $1-2^{1-N}|S_N|\geq 0$, which
implies that $|S_N| = \kappa_N 2^{N-1}$ for $0 \leq\kappa_N\leq 1$.

Let us then check for the range of values of $\kappa_N$ so that
the entanglement witnesses can detect entanglement for states
within the family and determine the minimal value of $N$. Before
doing this let us notice that it is enough to consider the states
from ${\cal F}$ for which $\alpha_N=0$. If $\alpha_N\neq 0$ then
we simply rotate $S_N$ as done previously for the operator $B_N$.

A straightforward computation gives us
$Tr(S_N\rho_N(\alpha_N=0))={1\over 1+N}(1-\kappa_N2^{N-1}+N)$,
which is negative if $\kappa_N> {1+N\over 2^{N-1}}$. Therefore,
it is enough to put $\kappa_N = 1$, in which case the
entanglement witnesses $S_N$ (strictly speaking rotated
entanglement witnesses) detect entanglement for all states from
the family ${\cal F}$, i.e., for $N\geq 4$.

It may be possible that the entanglement witnesses $S_N$ can be obtained from
some Bell inequality, perhaps using more than three settings of
measuring apparatus.

\section{Conclusions}
We have shown that the three-setting Bell inequalities derived in
Ref. \cite{ZUK-KASZ} are better for the detection of multipartite
bound entanglement for the family ${\cal F}$ presented in Ref.
\cite{DUR}. Application of the three-setting Bell inequalities
allows us to detect bound entanglement for seven qubits whereas
such a detection is possible for two-setting Bell inequalities
only for eight qubits and only for certain choice of the
parameter $\alpha_N$. We have also provided some numerical
evidence suggesting that bound entanglement of four qubits from
the family ${\cal F}$ cannot be detected in a Bell experiment in
which each observer uses two or three settings of the measuring
apparatus.

We have derived a family of entanglement witnesses that detect
bound entanglement for all members of the family ${\cal F}$,
i.e., for four and more qubits. The structure of these new
entanglement witnesses resembles the structure of entanglement
witnesses generated by two- and three-setting Bell inequalities
suggesting that it may be possible to find some Bell inequalities
that can detect bound entanglement for all members of the family
${\cal F}$.

\acknowledgments We acknowledge financial support provided under
the ASTAR Grant No. 012-104-0040.

\end{document}